\documentclass{emulateapj}  % for emulate style
\usepackage{epsf}
\usepackage{latexsym}  % for emulate style
\usepackage{epsfig}
\usepackage{natbib}

\def\msun{{\rm M}_\odot}
\def\mobs{M_{\rm obs}}
\def\mbias{M_{\rm bias}}

\def\pobs{P\left(M_{\rm obs}|M\right)}
\def\sigmalogm{\sigma_{\rm \ln M}}

\begin{document}

\bibstyle{aas}

\shorttitle{Non-Gaussian Scatter in Cluster Scaling Relations}
\shortauthors{Shaw et al.}

\author{Laurie D. Shaw, Gilbert P. Holder and Jonathan Dudley}
\affil{Department of Physics, McGill University, Montreal QC H3A 2T8}
\email{lds@physics.mcgill.ca}

\title{Non-Gaussian scatter in cluster scaling relations}

\begin{abstract}
We investigate the impact of non-Gaussian scatter in the cluster
mass-observable scaling relation on the mass and redshift distribution
of clusters detected by wide area surveys. We parameterize
non-Gaussian scatter by incorporating the third and forth moments
(skewness and kurtosis) into the distribution $\pobs$. We demonstrate
that for low scatter mass proxies the higher order moments do not
significantly affect the observed cluster mass and redshift
distributions. However, for high scatter mass indicators it is
necessary for the survey limiting mass threshold to be less than
$10^{14} h^{-1} \msun$ to prevent the skewness from having a
significant impact on the observed number counts, particularly at high
redshift. We also show that an unknown level of non-Gaussianity in the
scatter is equivalent to an additional uncertainty on the variance in
$\pobs$ and thus may limit the constraints that can be placed on the
dark energy equation of state parameter $w$. Furthermore, positive
skewness flattens the mass function at the high mass end, and so one
must also account for skewness in $\pobs$ when using the shape of the
mass function to constrain cluster scaling-relations.
\end{abstract}

\section{INTRODUCTION}

The evolution of the number density of galaxy clusters is a sensitive
cosmological probe \citep{bahcall:98,Eke:98b}.  As an indicator of the
expansion rate as a function of time, the galaxy cluster number
density is sensitive to the dark energy equation of state
\citep{haiman:00,weller:01}.  This provides a growth-based dark energy
test, an important complement to the distance-based tests that have
provided the most compelling evidence for dark energy to this point
\citep{perlmutter:99,schmidt:98}.  

Galaxy clusters can be selected by many diverse methods, including
(but not limited to) optical richness, X-ray thermal bremsstrahlung
flux, weak lensing shear and the Sunyaev-Zel'dovich (SZ) effect. The
key challenge for using galaxy clusters as precise cosmological probes
is in understanding how to relate observables to a quantity that can
be well predicted by theory, for example, mass. The ultimate goal is to
produce theoretical predictions of the distributions of observables as
a function of redshift and cosmological parameters. Short of this, one
approach is to theoretically model the evolution of number density as
a function of mass, and then estimate the mapping between observables
and mass in order to predict the observed evolution.  This mapping can
either be estimated from theoretical considerations or be determined
directly from the data, assuming some regularity in the mapping
\citep{Majumdar:03, hu:03, Lima:04, Lima:05}.

It is important to understand the statistics of the relevant
mass-observable scaling relation.  Although clusters may follow a mean
relation, individual clusters will deviate from it. If the level of
scatter around the mean relation is not small, the shape and amplitude
of the observed mass function can change significantly. At cluster
scales the mass-function is a steeply declining function of
mass. Therefore a larger number of low mass clusters will scatter over
the detection threshold of the survey than those higher mass clusters
that scatter below it. The net increase in the total number of
clusters in the sample is thus dependent on the slope of the mass
function at the threshold mass, and the magnitude of the scatter
around the mass-observable scaling relation. If the latter is well
constrained then the measurement of the cluster mass-function is
actually improved statistically due to the reduction in shot
noise. However, in practice it is difficult to precisely measure the
scatter and theoretical estimates can vary substantially. Furthermore,
as we shall demonstrate, for large intrinsic scatter, the magnitude of
the higher order moments (skewness and kurtosis) can become
significant.

Previous work forecasting the constraints on cosmological parameters
that can be achieved by cluster surveys have assumed the scatter in
scaling relations to be lognormally distribution around the mean
relation (that is, normally distributed in the logarithm of the
mass). However, cosmological simulations and observations of large
samples of clusters have demonstrated deviations from lognormal
behaviour. In general, the cause of these deviations can be seperated
into two catagories; dynamical state and projection effects. 

The former refers to the impact of a sub-population of clusters that
are systematically offset from the mean scaling relation due to their
dynamical state. \citet{Shaw:06} and \citet{Evrard:08} demonstrated
using N-body simulations that dark matter halos undergoing a major
merger have a systematically higher dark matter velocity dispersion
($\sigma_{\rm DM}$) than their more relaxed counterparts (of the same
mass). \citet{Evrard:08} showed that the small fraction of interacting
halos cause positive skewness in the distribution of the residuals
around the $mass - \sigma_{\rm DM}$ relation. \citet{Stanek:09}
analyse the covariance of bulk cluster properties using a large sample
of halos extracted from hydrodynamical simulations. They find small
deviations from Gaussian scatter for some properties, most notably in
the mass-weighted temperature -- mass relation. \citet{Pratt:09}
analysed X-ray luminosity scaling relations using a sample of 31
nearby clusters observed by the XMM-Newton X-ray observatory. They
find that the scatter in the $L_x - Y_X$ relation (where $Y_X$ was
assumed to be a robust, low scatter proxy for cluster mass) was
significantly non-Gaussian due to the systematically above-average
luminosity of cooling core clusters in their sample. They also note
that the scatter becomes more Gaussian when the central regions of the
clusters are excluded in the luminosity measurements.

The second cause of non-Gaussianity in observed scaling relations is
confusion in cluster selection due to projection effects. Several
studies have demonstrated that lower mass, unresolved clusters, as
well as gas outwith cluster regions can contribute significantly to
the measured integrated SZ flux ($Y$), causing additional scatter in
the $Y-M$ relation and introducing a tail towards high flux in the
distribution of $Y$ at constant mass \citep{White:02, Holder:07,
  Hallman:07}. A similar effect is found for optically-selected
clusters. \citet{Cohn:07} demonstrated using mock galaxy catalogues
that a non-negligable fraction of clusters (10\% at z = 0.4 to 22\% at
z = 1) identified using the red-sequence are `blends' -- cluster
candidates in which a large number of different halos have contributed
galaxies. Blends thus cause a tail towards high richness in the
distribution of optical richness at fixed mass.

In this work, we relax the assumption of lognormal scatter and
investigate the effect of non-Gaussian scatter around the mean
mass-observable scaling relation on the observed mass and redshift
distribution of clusters. Specifically, we quantify the impact of
non-zero skewness and kurtosis -- the third and fourth standardized
moments of a generalized probability distribution -- as perturbations
to the purely Gaussian case.

Throughout this paper we make a distinction between {\it true} mass
$M$ and {\it observed} mass $\mobs$. The former is the actual cluster
mass, as defined by a spherical overdensity, $\Delta$, and measured in
cosmological simulations of structure formation \citep{Jenkins:01,
  Warren:06, Tinker:08}. $\mobs$ is the cluster mass that would be
inferred from observations through application of a scaling relation
(e.g. $Y-M$, $L_x - M$, etc), or via self-calibration of the mass
function \citep{Lima:05, Cunha:09}.

\section{Scatter in the Mass-Observable Scaling Relation}
\label{sec:scatter}

The predicted redshift distribution of clusters observed by a given
survey is given by
\begin{equation}
\frac{dN}{dz} = \Delta \Omega \frac{dV}{dzd\Omega}(z)\int_0^{\infty}
\frac{dM}{M}\frac{d\bar n}{d\ln (M)} f(M,z)\;,
\label{eqn:massfunc}
\end{equation}
where $M$ is cluster mass and $d\bar n/d\ln M$ is the mean comoving
number density of clusters (the mass function).  The function
$f(M_{\rm lim},z)$ represents the survey selection function which
accounts for the limiting mass of the survey, $M_{\rm lim}$ defined by
some threshold in in the mass observable (e.g. optical richness, X-ray
luminosity, integrated SZ flux), and the statistics of the mapping
between $\mobs$ and $M$.  In the limit of perfect (zero scatter) mass
measurements, this is simply a step function at the limiting mass of
the survey.

If one assumes lognormal scatter around the mean scaling relation
(Gaussian scatter in $\ln M $) then the probability
$P\left(\mobs|M\right)$ of observing the mass $\mobs$ given the
``true'' underlying mass $M$, is
\begin{equation}
\pobs=\frac{1}{\sqrt{2\pi\sigma_{\ln M}^2}}\exp\left[-x^2\left(\mobs\right)\right]\;,
\label{eqn:pobs}
\end{equation}
with
\begin{equation}
x\left(\mobs\right)\equiv\frac{\ln M_{\rm obs} - \ln M -\ln
  \mbias}{\sqrt{2\sigma_{\rm \ln M}^2}}\;.
\label{eqn:massdist}
\end{equation}
This parameterization allows for redshift dependent scatter
$\sigmalogm$ and bias $\mbias$ in the mass-observable scaling relation
\citep{Lima:05, Cunha:09}. For simplicity, we henceforth ignore the
bias term and concentrate on the impact of scatter alone.  The
distribution of observed cluster masses is just a convolution of the true mass
function with $\pobs$,
\begin{equation}
\frac{d\bar n}{d \ln \mobs} = \int_0^\infty \frac{d\bar n}{d\ln M} \pobs d\ln M \;.
\label{eq:dndmobs}
\end{equation}
Plugging in Eqn. \ref{eqn:pobs} and assuming an intrinsic power law
distribution in mass, $d\bar n/d\ln (M) \propto M^{-\alpha}$, it is
straightforward to show that the observed mass distribution is
\begin{equation}
\frac{d\bar n}{d \ln \mobs} = \left(\frac{d\bar n}{d\ln M
}\right)_{\circ} e^{(\alpha^2\sigmalogm^2/2)} \;,
\label{eqn:gauss_scatter}
\end{equation}
where $\circ$ denotes the true mass distribution evaluated at $\mobs$
by applying the mean $M - \mobs$ scaling relation.

The extent of the deviation of the observed mass function from the
true mass function is clearly controlled by $\alpha \sigmalogm$, the
product of the standard deviation of the distribution $\pobs$ with the
slope of the mass function, $\alpha$. Evidently, a constant $\alpha
\sigmalogm$ results in an observed mass function that has a constant
and positive vertical offset from the true mass function. For more
realistic mass functions \citep{Jenkins:01, Tinker:08} the
slope $\alpha$ is approximately 1 at the group mass scale and
increases with increasing mass and redshift, exponentially so at very
high mass and redshift. The impact of scatter on the observed mass
function is thus significantly greater at high masses/redshifts (see
Section \ref{sec:surveys}).

Above a limiting threshold in $\mobs$, scatter in a cluster scaling
relation causes a net increase in the number of detected clusters.  An
unknown amount of scatter in the scaling relation thus degrades the
cosmological constraints that can be obtained from cluster number
counts as one must also marginalise over $\sigmalogm$. \citet{Lima:05}
demonstrated that for a survey with a fixed limiting mass of
$10^{14.2} h^{-1} \msun$ and $\sigma_{\ln M}^2 = 0.25^2$, a $1\sigma$
uncertainty of $0.25^2$ on $\sigma^2_{\ln M}$ would produce a 10\%
uncertainty in the number counts at $z = 0.5$, a 20\% uncertainty at
$z = 1$, and a 50\% uncertainty at $z = 2$.

\subsection{Non-Gaussian Scatter}

We now determine the impact on the observed mass function of
non-Gaussian scatter in the mass-observable relation. We proceed by
using the Edgeworth series to approximate a non-Gaussian $\pobs$
\citep[e.g.][]{Bernardeau:95, Blinnikov:98}, thus taking it to be a
perturbation to the Gaussian case. It is parameterised by the third
and fourth moments, the skewness ($\gamma$) and kurtosis ($\kappa$),
of the probability distribution $\pobs$. The Edgeworth expansion is
particularly useful for convolutions when expressed as a series of
derivatives of Gaussians,
\begin{equation}
\pobs \approx G(x) - \frac{\gamma}{6}\frac{d^3 G}{dx^3} +
\frac{\kappa}{24}\frac{d^4 G}{dx^4} +  \frac{\gamma^2}{72}\frac{d^6 G}{dx^6}\; 
\label{eq:pobs_ng}
\end{equation}
where the skewness, $\gamma$ is defined as
\begin{equation}
\gamma = \frac{\left<\left(\mobs-M\right)^3\right>}{\sigma^3}
\end{equation}
and the kurtosis, $\kappa$ as
\begin{equation}
\kappa = \frac{\left<\left(\mobs-M\right)^4\right>}{\sigma^4} -3\;,
\end{equation} 
and $G(x)$ is a Gaussian distribution in $x$ (equal to $\pobs$ in
Eq. \ref{eqn:pobs}).

By plugging Equation \ref{eq:pobs_ng} into Equation \ref{eq:dndmobs}
and doing some integration by parts, the observed mass function can be
calculated for any true mass function,
\begin{eqnarray}
{d\bar n \over d \ln M_{obs}} &=&  \int {d\bar n \over dx } G(x) dx -
{\gamma \over 6} \int {d^3 \over dx^3}\Bigl({d\bar n\over dx }\Bigr ) G(x) dx  \nonumber\\
&&\qquad+ {\kappa \over 24} \int {d^4 \over dx^4}\Bigl( {d\bar n \over dx }\Bigr )G(x) dx  \nonumber\\
&&\qquad+ {\gamma^2 \over 72} \int {d^6 \over dx^6}\Bigl({d\bar n\over dx }\Bigr )G(x) dx + ... 
\label{eqn:non_guass_general}
\end{eqnarray}
Assuming again an intrinsic power-law distribution of mass with slope
$\alpha$, $d\bar n/d\ln M_{\rm true} = M^{-\alpha}$, it is straightforward
to show that the observed mass distribution is now
\begin{eqnarray} 
{d\bar n \over d \ln \mobs} &=&  \Bigl({d\bar n \over d\ln M} \Bigr)_\circ e^{\alpha^2\sigma^2/2} \times \nonumber\\
&&\Bigl [ 1 + {\alpha^3 \sigma^3 \over 6}\gamma  + 
 {\alpha^4 \sigma^4 \over 24 }\kappa + {\alpha^6 \sigma^6 \over 72} \gamma^2 \Bigr] \;, 
\label{eqn:mod_powerlaw}
\end{eqnarray}
where $\circ$ denotes the true mass distribution evaluated at $\mobs$.

The relevant parameter here is again clearly $\alpha \sigmalogm$.  If
this parameter is large, there are two important effects: the number
of objects at any given mass scale is increased substantially (as for
the Gaussian case), and the higher order moments of the distribution,
$\gamma$ and $\kappa$, become important. The transition at which the
latter occurs is clearly when $\alpha \sigmalogm$ becomes greater than
unity. For example, if $\alpha \sigmalogm = 1$, then $\gamma =
1$ produces an 17\% increase on the number counts compared to the purely
Gaussian case at any given mass scale. Assuming $\kappa = 1$ provides
an additional 4\% correction. Note that Equation
\ref{eqn:mod_powerlaw} demonstrates that both (positive) $\gamma$ and
$\kappa$ cause an up-scattering of clusters. For skewness this is due
to the tail towards large mass that increases the probability of a
cluster of having $\mobs >> M$. For kurtosis, the up-scattering is due
to the wider wings of the distribution.

\section{Implications of non-Gaussian scatter for SZ and Optical Cluster Surveys}
\label{sec:surveys}

\begin{figure}
\plotone{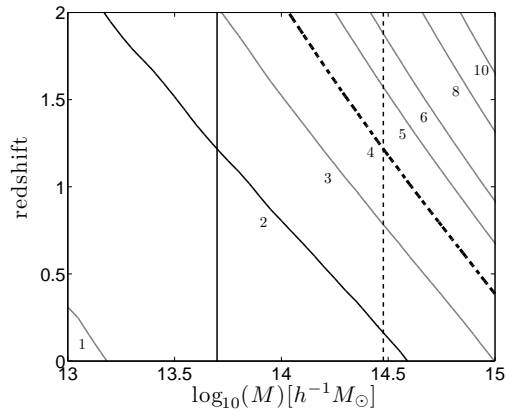}
\caption{Contours of constant slope, $\alpha$ of the halo mass
  function, where $d\bar n/d\ln (M) \propto M^{-\alpha}$, as a function of
  mass and redshift for our fiducial cosmology. The solid black
  contour emphasizes the critical curve $M_c(z)$ (see text) for a
  DES-like optical survey ($\sigmalogm = 0.5$), the black dashed
  contour shows $M_c(z)$ for an SPT-like SZ survey ($\sigmalogm =
  0.25$). The black solid and dashed vertical lines show the assumed
  limiting masses for the DES and SPT surveys, respectively.}
\label{fig:mf_alpha}
\end{figure}

We now evaluate the impact of non-Gaussian scatter in the
mass-observable scaling relation on the predicted mass and redshift
distribution of clusters observed by a South Pole Telescope-like SZ
survey and a Dark Energy Survey-like optical survey. For the SZ survey
we assume that the scatter in the mass-SZ flux relation is
$\sigma_{\rm SZ} = 0.25$ \citep{Shaw:08} and the limiting mass is
$3\times10^{14} h^{-1} \msun$. For the optical survey we assume the
scatter in the optical richness - mass relation is $\sigma_{\rm opt} =
0.5$, in keeping with the results of \citet{Rykoff:08a, Rykoff:08b,
  Becker:07, Rozo:09, Rozo:08b}, and a constant limiting mass of
$5\times 10^{13} h^{-1} \msun$. Note that \citet{Becker:07} found that
the scatter in the richness-mass relation is 0.5-0.75, and we have
taken the value at the lower end of this range.

We have demonstrated that the impact of the higher order moments
$\gamma$ and $\kappa$ become significant when $\sigmalogm \alpha$, the
product of the slope of the mass function and the standard deviation
of $\pobs$, becomes greater than unity. Under the simplifying
assumption that $\sigmalogm$ is independent of mass and redshift for a
given observable, the higher order moments thus become relevant when
the effective slope of the mass function exceeds the threshold value,
$\alpha_c > 1/\sigmalogm$. This critical slope is reached above a
redshift (and cosmology) dependent mass threshold $M_c(z)$.  For SPT,
$\alpha_{\rm c,SZ} = 1/0.25 = 4$, for DES $\alpha_{\rm c,opt} = 2$.

In Figure \ref{fig:mf_alpha} we plot the slope of the mass function as
a function of mass and redshift. For this, we adopt the mass function
of \citet{Tinker:08}, and assuming $d\bar n/d\ln(M) \propto
M^{-\alpha}$ at any given mass and redshift, calculate $\alpha$ over a
wide range of mass and redshift. The mass function is defined in terms
of $M_{200}$, where the subscipt denotes the mass enclosed in the
region of spherical overdensity of 200 times the {\it mean} density of
the Universe (at the relevant redshift). For our fiducial cosmology we
assume parameters consistent with the WMAP 5-year results
\citep[$\Omega_{\rm M} = 0.27$, $\sigma_8 = 0.8$,][]{Dunkley:09}. The
grey lines denote contours of constant $\alpha$, from 10 (top right
corner) to 1 (bottom left corner). The thick black solid and dashed
contours denote $\alpha = 2$ and 4, the critical slopes $\alpha_c$ for
DES and SPT, respectively. These contours thus give $M_c(z)$ for each
survey. The vertical solid and dashed lines represent the limiting
mass for each survey.

The Figure demonstrates that for the SZ survey, $\alpha \sigmalogm$
becomes greater than unity above $8\times10^{14} h^{-1} \msun$ at z=
0.5, and above $4\times10^{14} h^{-1} \msun$ at z = 1. Given the
scarcity of objects above these thresholds, it is clear that $\gamma$
and $\kappa$ will not significantly impact the SPT cluster number
counts.

For the optical survey, $M_c(z)$ is $3\times10^{14} h^{-1} \msun$ at z
= 0.2, $1.7\times10^{14} h^{-1}\msun$ at z = 0.5 and
$5.25\times10^{13} h^{-1}\msun$ at z = 1.2. As $M_c(z)$ remains
greater than the expected limiting mass for DES, $\gamma$ and $\kappa$
will not strongly effect the total number of detected
clusters. However, a sizeable number of clusters will be detected in
mass bins that exceed $M_c(z)$. Non-Gaussian scatter in the
mass-richness relation may therefore cause a detectable redistribution
of clusters in mass bins greater than $M_c(z)$,

\begin{figure}
\plotone{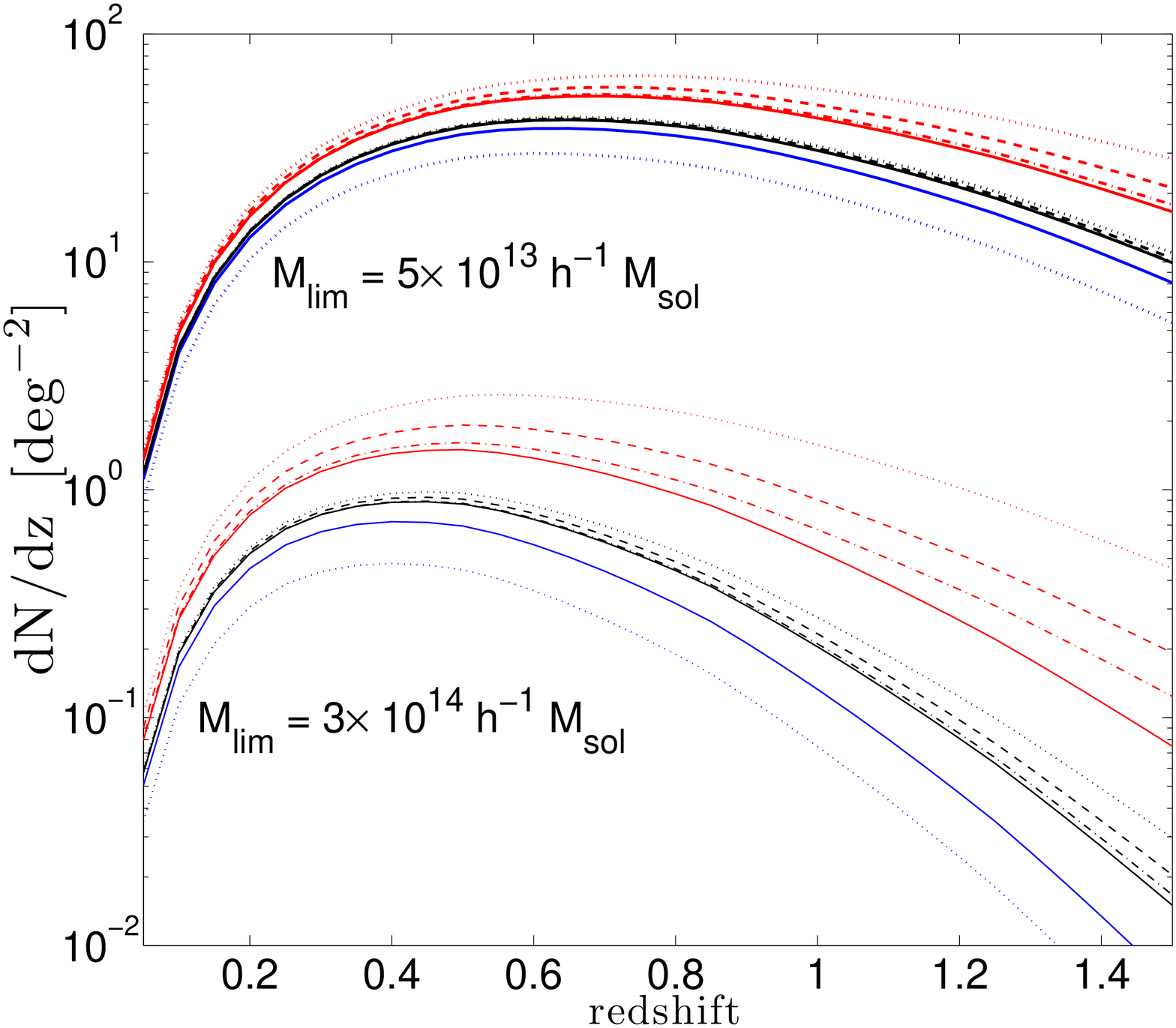}
\plotone{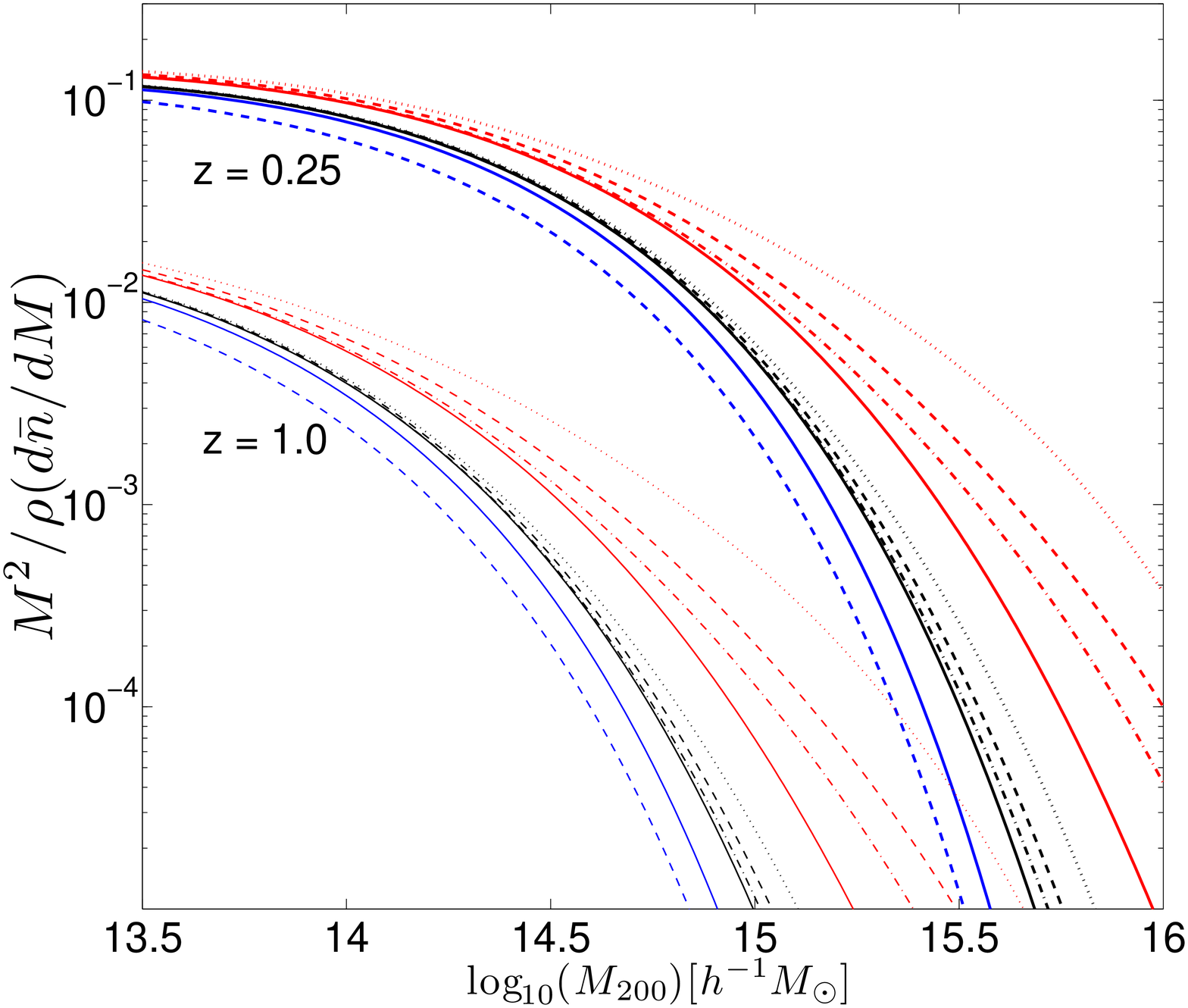}
\caption{({\it Upper}) The observed redshift distribution of clusters,
  $dN/dz$ above a limiting observed mass threshold of $5\times 10^{13}
  h^{-1} \msun$ and $3\times 10^{14} h^{-1} \msun$. ({\it Lower}) The
  observed mass distribution of clusters $\mobs^2/\rho_m (d\bar n/d
  \mobs)$ at z = 0.25 and z = 1. In each plot, the line colors
  represent the value of $\sigmalogm$; 0 (blue), 0.25 (black) and 0.5
  (red). The solid lines represent $\gamma = \kappa = 0$, the dashed
  $\gamma = 0.5, \kappa = 0$, dot-dashed $\kappa = 0.5$ $(\gamma = 0)$
  and the dotted line $\gamma = 1$, $\kappa = 1$ (in our fiducial
  cosmology). The blue dotted line shows the results for a WMAP3
  cosmology, where $\sigma_8 = 0.77$, $\Omega_{\rm M} = 0.24$, $H_0 =
  73 {\rm km/s/Mpc}$, and we take $\sigmalogm = 0$.}
\label{fig:massdist}
\end{figure}

In Figure \ref{fig:massdist} we demonstrate the impact of skewness and
kurtosis in $\pobs$ on the observed redshift and mass distribution of
clusters. In both panels, the colors of each line represent the value
of $\sigmalogm$; 0 (blue), 0.25 (black) and 0.5 (red). The solid lines
represent (in our fiducial WMAP5 cosmology) $\gamma = \kappa = 0$,
dashed $\gamma = 0.5, \kappa = 0$, dot-dashed $\kappa = 0.5$ $(\gamma
= 0)$ and dotted $\gamma = 1$, $\kappa = 1$ (as a more extreme
scenario). The blue dotted line shows the results for a WMAP 3-year
cosmology, where $\sigma_8 = 0.77$, $\Omega_{\rm M} = 0.24$, $H_0 = 73
 {\rm km/s/Mpc}$, and $\sigmalogm = 0$ \citep{Spergel:07}.

In the upper panel we plot the redshift distribution of clusters
$dN/dz$ using our two $\mobs$ limits; $5\times 10^{13} h^{-1} \msun$
for DES (upper lines) and $3\times 10^{14} h^{-1} \msun$ for SPT
(lower lines). For the purposes of comparison, we have plotted the
results for $\sigmalogm = 0.25$ and $0.5$ for both limiting masses. As
predicted by Equation \ref{eqn:mod_powerlaw}, the impact of skewness
and kurtosis on the redshift distribution is dependent on the value of
$\sigmalogm$ and the local slope of the (true) mass function at the
limiting mass of the survey. Thus at higher redshifts the impact of
$\gamma$ and $\kappa$ becomes greater. The flattening of $dN/dz$ due
to the higher order momements is more pronounced than that due to an
increase in $\sigma_8$ or $\sigmalogm$.

For a threshold of $M_{\rm obs,lim} = 5\times 10^{13} h^{-1} \msun$
and $\sigmalogm = 0.5$ a skewness of $\gamma$ $(\kappa)= 0.5$
increases the mean number of clusters by 10\% (2\%) at $z = 0.75$. The
more extreme case of $\gamma = \kappa = 1$ produces a 24\%
increase. For the same scatter but with a limiting $\mobs$ of $3\times
10^{14} h^{-1} \msun$, $\gamma$ $(\kappa)= 0.5$ increases the number
of clusters at $z = 0.75$ by 42\% (14\%), and $\gamma = \kappa = 1$ by
218\%. It is clear that it is the low mass threshold of the optical
survey that prevents the higher order moments from strongly affecting
the number of detected clusters in each redshift bin. The increase in
$dN/dz$ (at z = 0.75) for $\gamma = 0.5$ is roughly equivalent to
assuming lognormal scatter but increasing $\sigmalogm$ by 22\% to 0.61
(for $M_{\rm obs,lim} = 3\times 10^{14} h^{-1} \msun$) and by 14\% to
0.57 ($M_{\rm obs,lim} = 5\times 10^{13} h^{-1} \msun$). For our
extreme case of $\gamma = \kappa = 1$, the results are equivalent to
increasing $\sigmalogm$ to 0.73 and 0.65 at each mass threshold,
respectively.

For the SZ survey ($\sigmalogm = 0.25$, $M_{\rm obs,lim} = 3\times
10^{14} h^{-1} \msun$), increasing $\gamma$ from 0 to 0.5 provides a
8\% increase in the number counts at z = 0.75. This is equivalent to
an increase in $\sigmalogm$ for purely lognormal scatter of 12\% (to
0.28). The same change in kurtosis provides less than a one percent
change in the number counts. The $\gamma = \kappa = 1$ case provides a
20\% increase in the number counts, equivalent to an increase of
$\sigmalogm$ by 28\% to 0.32.

In the lower panel of Figure \ref{fig:massdist} we plot the mass
distribution of detected clusters, $\mobs^2/\rho_{\rm m} (d\bar n/d
\mobs)$ (scaled to reduce the dynamic range, where $\rho_{\rm m}$ is
the mean matter density of the Universe at each redshift) at z = 0.25
and 1.0. The line types and colors represent the same values of
$\sigmalogm$, $\gamma$, and $\kappa$ as the upper plot. For integrated
SZ flux (for which $\sigmalogm = 0.25$), the higher-order moments do
not have a significant effect on the shape of the mass function. The
impact of $\gamma$ and $\kappa$ is more evident for high-scatter mass
indicators (red lines), flattening the mass distribution above
$10^{14.5} h^{-1} \msun$ at z = 0.25, and $10^{14} h^{-1} \msun$ at z
= 1.0. These masses roughly correspond to the predicted values of
$M_c(z)$ at these redshifts in Figure \ref{fig:mf_alpha} ($\alpha = 2$
contour).

Overall, Figure \ref{fig:massdist} demonstrates two points. A skewness
of $\gamma = 0.5$ has the same effect on dN/dz (at z = 0.75) as
increasing the variance $\sigmalogm^2$ by 0.075 for DES and by 0.016
for SPT. For the significantly more non-Gaussian distribution
parameterised by $\gamma = \kappa = 1$, the equivalent change in
$\sigmalogm^2$ is 0.17 for DES and 0.04 for SPT. \citet{Lima:05} show
that a prior of $\sigma(\sigmalogm^2) = 0.01$ is necessary to ensure
that uncertainty on the level of scatter does not degrade the
constraints that can be placed on the dark energy equation of state
parameter $w$ (see their Figure 6). Therefore, assuming $\gamma$ does
not greatly exceed 0.5, non-Gaussian scatter in the mass-SZ flux
scaling relation should not significantly degrade the cosmological
constraints that can be achieved by SPT. However, for surveys
utilizing a high-scatter mass proxy such as optical richness, an
unknown level of non-Gaussianity in $\pobs$ may limit the constraints
that can be placed on the dark energy equation of state via cluster
number counts.

Secondly, if one aims to use the shape of the mass function to
self-calibrate a survey to obtain information on the slope,
normalization and variance of the mass-observable scaling relation,
then care must be taken to ensure that the parameters are not biased
due to the increasing impact of the higher order moments on the shape
of the mass function towards high masses. One means of ensuring this
is to include only a single mass bin for $\mobs > M_c(z)$.

\section{Conclusion}

We have investigated the impact of non-Gaussian scatter in the
mass-observable scaling relation on the cluster mass function by
incorporating the third and forth moments (skewness and kurtosis) into
the probability distribution $\pobs$ via the Edgeworth expansion. We
have demonstrated that if $\alpha \sigmalogm$ -- the product of the
standard deviation in $\pobs$ and the slope of the mass function at
the limiting mass of a survey -- is greater than unity, then positive
skewness and kurtosis will increase the number of clusters detected
and flatten the high-mass end of the observed mass function.

For low scatter mass proxies like integrated SZ flux, higher order
moments do not significantly affect the mass and redshift distribution
of clusters. However, for surveys utilizing a high scatter mass proxy
such as optical richness, the limiting mass threshold must be less
than $10^{14} h^{-1} \msun$ to ensure that the skewness does not
significantly effect $dN/dz$, especially at high redshift. We have
also found that an unknown level of non-Gaussian scatter is roughly
equivalent to an additional uncertainty on the variance $\sigmalogm^2$
in $\pobs$ and thus may limit the contraints that can be placed on the
dark energy equation of state parameter $w$, especially for surveys
that use a high scatter mass proxy. Furthermore, if one wishes to use
the shape of the mass function in each redshift bin to self-calibrate
for cluster scaling relation parameters then it will be necessary to
account for non-Gaussian scatter on the shape of the mass function by
marginalising over the skewness and kurtosis parameters, $\gamma$ and
$\kappa$ in addition to the the variance $\sigmalogm^2$, slope and
normalisation in each redshift bin.

We note that the values of $\gamma$ and $\kappa$ used in the examples
given in this work were chosen arbitrarily (and to be small enough to
ensure that the Edgeworth expansion remains an appropriate
approximation to a non-Gaussian $P(\mobs|M)$). Large-volume
simulations and mock-galaxy catalogues containing several thousands of
clusters will be necessary to obtain better motivated predictions of
$\gamma$ and $\kappa$ for optical, SZ (and X-ray) surveys.  Measuring
$\gamma$ and $\kappa$ would require a large observational sample with
precisely measured masses (for example, using X-ray spectroscopic
temperatures). However, the higher order statistics are strongly
influenced by rare, outlying objects (e.g. major mergers), or the
impact of observational selection effects such as cluster-cluster
confusion. For this reason, it is unlikely that it will be possible to
place tight priors on these parameters.

\section{ACKNOWLEDGMENTS}
This work is supported by NSERC through the Discovery Grant Awards to
GPH. GPH would also like to acknowledge support from the Canadian
Institute for Advanced Research and the Canada Research Chairs
Program. We would like to thank Jochen Weller, Gus Evrard and Eduardo
Rozo for useful discussions.

%\def\apj{Ap.\ J.}  
%\bibliography{../biblist.bib}

\begin{thebibliography}{32}
\expandafter\ifx\csname natexlab\endcsname\relax\def\natexlab#1{#1}\fi

\bibitem[{Bahcall \& Fan(1998)}]{bahcall:98}
Bahcall, N. \& Fan, X. 1998, \apj, 504, 1

\bibitem[{{Becker} {et~al.}(2007){Becker}, {McKay}, {Koester}, {Wechsler},
  {Rozo}, {Evrard}, {Johnston}, {Sheldon}, {Annis}, {Lau}, {Nichol}, \&
  {Miller}}]{Becker:07}
{Becker}, M.~R., {McKay}, T.~A., {Koester}, B., {Wechsler}, R.~H., {Rozo}, E.,
  {Evrard}, A., {Johnston}, D., {Sheldon}, E., {Annis}, J., {Lau}, E.,
  {Nichol}, R., \& {Miller}, C. 2007, \apj, 669, 905

\bibitem[{{Bernardeau} \& {Kofman}(1995)}]{Bernardeau:95}
{Bernardeau}, F. \& {Kofman}, L. 1995, \apj, 443, 479

\bibitem[{{Blinnikov} \& {Moessner}(1998)}]{Blinnikov:98}
{Blinnikov}, S. \& {Moessner}, R. 1998, \aaps, 130, 193

\bibitem[{{Cohn} {et~al.}(2007){Cohn}, {Evrard}, {White}, {Croton}, \&
  {Ellingson}}]{Cohn:07}
{Cohn}, J.~D., {Evrard}, A.~E., {White}, M., {Croton}, D., \& {Ellingson}, E.
  2007, \mnras, 382, 1738

\bibitem[{{Cunha}(2009)}]{Cunha:09}
{Cunha}, C. 2009, \prd, 79, 063009

\bibitem[{{Dunkley} {et~al.}(2009){Dunkley}, {Komatsu}, {Nolta}, {Spergel},
  {Larson}, {Hinshaw}, {Page}, {Bennett}, {Gold}, {Jarosik}, {Weiland},
  {Halpern}, {Hill}, {Kogut}, {Limon}, {Meyer}, {Tucker}, {Wollack}, \&
  {Wright}}]{Dunkley:09}
{Dunkley}, J., {Komatsu}, E., {Nolta}, M.~R., {Spergel}, D.~N., {Larson}, D.,
  {Hinshaw}, G., {Page}, L., {Bennett}, C.~L., {Gold}, B., {Jarosik}, N.,
  {Weiland}, J.~L., {Halpern}, M., {Hill}, R.~S., {Kogut}, A., {Limon}, M.,
  {Meyer}, S.~S., {Tucker}, G.~S., {Wollack}, E., \& {Wright}, E.~L. 2009,
  \apjs, 180, 306

\bibitem[{{Eke} {et~al.}(1998){Eke}, {Cole}, {Frenk}, \& {Patrick
  Henry}}]{Eke:98b}
{Eke}, V.~R., {Cole}, S., {Frenk}, C.~S., \& {Patrick Henry}, J. 1998, \mnras,
  298, 1145

\bibitem[{{Evrard} {et~al.}(2008){Evrard}, {Bialek}, {Busha}, {White}, {Habib},
  {Heitmann}, {Warren}, {Rasia}, {Tormen}, {Moscardini}, {Power}, {Jenkins},
  {Gao}, {Frenk}, {Springel}, {White}, \& {Diemand}}]{Evrard:08}
{Evrard}, A.~E., {Bialek}, J., {Busha}, M., {White}, M., {Habib}, S.,
  {Heitmann}, K., {Warren}, M., {Rasia}, E., {Tormen}, G., {Moscardini}, L.,
  {Power}, C., {Jenkins}, A.~R., {Gao}, L., {Frenk}, C.~S., {Springel}, V.,
  {White}, S.~D.~M., \& {Diemand}, J. 2008, \apj, 672, 122

\bibitem[{Haiman {et~al.}(2001)Haiman, Mohr, \& Holder}]{haiman:00}
Haiman, Z., Mohr, J.~J., \& Holder, G.~P. 2001, \apj, 553, 545

\bibitem[{{Hallman} {et~al.}(2007){Hallman}, {O'Shea}, {Burns}, {Norman},
  {Harkness}, \& {Wagner}}]{Hallman:07}
{Hallman}, E.~J., {O'Shea}, B.~W., {Burns}, J.~O., {Norman}, M.~L., {Harkness},
  R., \& {Wagner}, R. 2007, \apj, 671, 27

\bibitem[{{Holder} {et~al.}(2007){Holder}, {McCarthy}, \& {Babul}}]{Holder:07}
{Holder}, G.~P., {McCarthy}, I.~G., \& {Babul}, A. 2007, \mnras, 382, 1697

\bibitem[{{Hu}(2003)}]{hu:03}
{Hu}, W. 2003, \prd, 67, 081304

\bibitem[{{Jenkins} {et~al.}(2001){Jenkins}, {Frenk}, {White}, {Colberg},
  {Cole}, {Evrard}, {Couchman}, \& {Yoshida}}]{Jenkins:01}
{Jenkins}, A., {Frenk}, C.~S., {White}, S.~D.~M., {Colberg}, J.~M., {Cole}, S.,
  {Evrard}, A.~E., {Couchman}, H.~M.~P., \& {Yoshida}, N. 2001, \mnras, 321,
  372

\bibitem[{{Lima} \& {Hu}(2004)}]{Lima:04}
{Lima}, M. \& {Hu}, W. 2004, \prd, 70, 043504

\bibitem[{{Lima} \& {Hu}(2005)}]{Lima:05}
---. 2005, \prd, 72, 043006

\bibitem[{{Majumdar} \& {Mohr}(2003)}]{Majumdar:03}
{Majumdar}, S. \& {Mohr}, J.~J. 2003, \apj, 585, 603

\bibitem[{Perlmutter {et~al.}(1999)Perlmutter, Aldering, Goldhaber, Knop,
  Nugent, Castro, Deustua, Fabbro, Goobar, Groom, Hook, Kim, Kim, Lee, Nunes,
  Pain, Pennypacker, Quimby, Lidman, Ellis, Irwin, McMahon, Ruiz-Lapuente,
  Walton, Schaefer, Boyle, Filippenko, Matheson, Fruchter, Panagia, Newberg, \&
  Couch}]{perlmutter:99}
Perlmutter, S., Aldering, G., Goldhaber, G., Knop, R., Nugent, P., Castro, P.,
  Deustua, S., Fabbro, S., Goobar, A., Groom, D.~E., Hook, I.~M., Kim, A.~G.,
  Kim, M., Lee, J., Nunes, N., Pain, R., Pennypacker, C., Quimby, R., Lidman,
  C., Ellis, R., Irwin, M., McMahon, R., Ruiz-Lapuente, P., Walton, N.,
  Schaefer, B., Boyle, B., Filippenko, A., Matheson, T., Fruchter, A., Panagia,
  N., Newberg, H. J.~M., \& Couch, W. 1999, \apj, 517, 565

\bibitem[{{Pratt} {et~al.}(2009){Pratt}, {Croston}, {Arnaud}, \&
  {B{\"o}hringer}}]{Pratt:09}
{Pratt}, G.~W., {Croston}, J.~H., {Arnaud}, M., \& {B{\"o}hringer}, H. 2009,
  \aap, 498, 361

\bibitem[{{Rozo} {et~al.}(2009){Rozo}, {Rykoff}, {Evrard}, {Becker}, {McKay},
  {Wechsler}, {Koester}, {Hao}, {Hansen}, {Sheldon}, {Johnston}, {Annis}, \&
  {Frieman}}]{Rozo:09}
{Rozo}, E., {Rykoff}, E.~S., {Evrard}, A., {Becker}, M., {McKay}, T.,
  {Wechsler}, R.~H., {Koester}, B.~P., {Hao}, J., {Hansen}, S., {Sheldon}, E.,
  {Johnston}, D., {Annis}, J., \& {Frieman}, J. 2009, \apj, 699, 768

\bibitem[{{Rozo} {et~al.}(2008){Rozo}, {Rykoff}, {Koester}, {McKay}, {Hao},
  {Evrard}, {Wechsler}, {Hansen}, {Sheldon}, {Johnston}, {Becker}, {Annis},
  {Bleem}, \& {Scranton}}]{Rozo:08b}
{Rozo}, E., {Rykoff}, E.~S., {Koester}, B.~P., {McKay}, T., {Hao}, J.,
  {Evrard}, A., {Wechsler}, R.~H., {Hansen}, S., {Sheldon}, E., {Johnston}, D.,
  {Becker}, M., {Annis}, J., {Bleem}, L., \& {Scranton}, R. 2008, ArXiv
  e-prints

\bibitem[{{Rykoff} {et~al.}(2008{\natexlab{a}}){Rykoff}, {Evrard}, {McKay},
  {Becker}, {Johnston}, {Koester}, {Nord}, {Rozo}, {Sheldon}, {Stanek}, \&
  {Wechsler}}]{Rykoff:08a}
{Rykoff}, E.~S., {Evrard}, A.~E., {McKay}, T.~A., {Becker}, M.~R., {Johnston},
  D.~E., {Koester}, B.~P., {Nord}, B., {Rozo}, E., {Sheldon}, E.~S., {Stanek},
  R., \& {Wechsler}, R.~H. 2008{\natexlab{a}}, \mnras, 387, L28

\bibitem[{{Rykoff} {et~al.}(2008{\natexlab{b}}){Rykoff}, {McKay}, {Becker},
  {Evrard}, {Johnston}, {Koester}, {Rozo}, {Sheldon}, \&
  {Wechsler}}]{Rykoff:08b}
{Rykoff}, E.~S., {McKay}, T.~A., {Becker}, M.~R., {Evrard}, A., {Johnston},
  D.~E., {Koester}, B.~P., {Rozo}, E., {Sheldon}, E.~S., \& {Wechsler}, R.~H.
  2008{\natexlab{b}}, \apj, 675, 1106

\bibitem[{{Schmidt} {et~al.}(1998){Schmidt}, {Suntzeff}, {Phillips},
  {Schommer}, {Clocchiatti}, {Kirshner}, {Garnavich}, {Challis}, {Leibundgut},
  {Spyromilio}, {Riess}, {Filippenko}, {Hamuy}, {Smith}, {Hogan}, {Stubbs},
  {Diercks}, {Reiss}, {Gilliland}, {Tonry}, {Maza}, {Dressler}, {Walsh}, \&
  {Ciardullo}}]{schmidt:98}
{Schmidt}, B.~P., {Suntzeff}, N.~B., {Phillips}, M.~M., {Schommer}, R.~A.,
  {Clocchiatti}, A., {Kirshner}, R.~P., {Garnavich}, P., {Challis}, P.,
  {Leibundgut}, B., {Spyromilio}, J., {Riess}, A.~G., {Filippenko}, A.~V.,
  {Hamuy}, M., {Smith}, R.~C., {Hogan}, C., {Stubbs}, C., {Diercks}, A.,
  {Reiss}, D., {Gilliland}, R., {Tonry}, J., {Maza}, J.~e., {Dressler}, A.,
  {Walsh}, J., \& {Ciardullo}, R. 1998, \apj, 507, 46

\bibitem[{{Shaw} {et~al.}(2008){Shaw}, {Holder}, \& {Bode}}]{Shaw:08}
{Shaw}, L.~D., {Holder}, G.~P., \& {Bode}, P. 2008, \apj, 686, 206

\bibitem[{{Shaw} {et~al.}(2006){Shaw}, {Weller}, {Ostriker}, \&
  {Bode}}]{Shaw:06}
{Shaw}, L.~D., {Weller}, J., {Ostriker}, J.~P., \& {Bode}, P. 2006, \apj, 646,
  815

\bibitem[{{Spergel} {et~al.}(2007){Spergel}, {Bean}, {Dor{\'e}}, {Nolta},
  {Bennett}, {Dunkley}, {Hinshaw}, {Jarosik}, {Komatsu}, {Page}, {Peiris},
  {Verde}, {Halpern}, {Hill}, {Kogut}, {Limon}, {Meyer}, {Odegard}, {Tucker},
  {Weiland}, {Wollack}, \& {Wright}}]{Spergel:07}
{Spergel}, D.~N., {Bean}, R., {Dor{\'e}}, O., {Nolta}, M.~R., {Bennett}, C.~L.,
  {Dunkley}, J., {Hinshaw}, G., {Jarosik}, N., {Komatsu}, E., {Page}, L.,
  {Peiris}, H.~V., {Verde}, L., {Halpern}, M., {Hill}, R.~S., {Kogut}, A.,
  {Limon}, M., {Meyer}, S.~S., {Odegard}, N., {Tucker}, G.~S., {Weiland},
  J.~L., {Wollack}, E., \& {Wright}, E.~L. 2007, \apjs, 170, 377

\bibitem[{{Stanek} {et~al.}(2009){Stanek}, {Rasia}, {Evrard}, {Pearce}, \&
  {Gazzola}}]{Stanek:09}
{Stanek}, R., {Rasia}, E., {Evrard}, A.~E., {Pearce}, F., \& {Gazzola}, L.
  2009, in preparation

\bibitem[{{Tinker} {et~al.}(2008){Tinker}, {Kravtsov}, {Klypin}, {Abazajian},
  {Warren}, {Yepes}, {Gottl{\"o}ber}, \& {Holz}}]{Tinker:08}
{Tinker}, J., {Kravtsov}, A.~V., {Klypin}, A., {Abazajian}, K., {Warren}, M.,
  {Yepes}, G., {Gottl{\"o}ber}, S., \& {Holz}, D.~E. 2008, \apj, 688, 709

\bibitem[{{Warren} {et~al.}(2006){Warren}, {Abazajian}, {Holz}, \&
  {Teodoro}}]{Warren:06}
{Warren}, M.~S., {Abazajian}, K., {Holz}, D.~E., \& {Teodoro}, L. 2006, \apj,
  646, 881

\bibitem[{{Weller} {et~al.}(2001){Weller}, {Battye}, \& {Kneissl}}]{weller:01}
{Weller}, J., {Battye}, R., \& {Kneissl}, R. 2001, \prl, 88, 231301

\bibitem[{{White} {et~al.}(2002){White}, {Hernquist}, \& {Springel}}]{White:02}
{White}, M., {Hernquist}, L., \& {Springel}, V. 2002, \apj, 579, 16

\end{thebibliography}
%\bibliographystyle{/Users/laurieshaw/Documents/Research/papers/astronat/apj/apj}

\end{document}